\documentclass[%
 reprint,
nobibnotes,
 amsmath,amssymb,
 aps,
]{revtex4-1}

\usepackage{graphicx}
\usepackage{dcolumn}
\usepackage{bm}

\begin{document}

\preprint{APS/123-QED}

\title{Breaking the temporal resolution limit by superoscillating optical beats}

\author{Yaniv Eliezer $\ddagger$}
\author{Liran Hareli $\ddagger$}
\author{Lilya Lobachinsky}
\author{Sahar Froim}
\author{Alon Bahabad}
	 \email{alonb@eng.tau.ac.il}

\affiliation{ Department of Physical Electronics, School of Electrical Engineering, Fleischman Faculty of Engineering, Tel-Aviv University, Tel-Aviv 69978, Israel }

\date{\today}

\begin{abstract}
	Band-limited functions can oscillate locally at an arbitrarily fast rate through an interference phenomenon known as superoscillations. Using an optical pulse with a superoscillatory envelope we experimentally break the temporal Fourier-transform limit having a temporal feature which is approximately three times shorter than the duration of a transform-limited Gaussian pulse having a comparable bandwidth while maintaining $29.5\%$ visibility. Numerical simulations demonstrate the ability of such signals to achieve temporal super-resolution.
\end{abstract}

\maketitle



	{\emph{Introduction}. The ability to manipulate the waveform of optical pulses is essential for numerous applications \cite{keller2003recent} such as spectroscopy and coherent control \cite{goswami2003optical, silberberg2009quantum}, 
	metrology \cite{tanabe2002spatiotemporal} , microscopy \cite{bardeen1999effect} and optical communications \cite{sardesai1998femtosecond}.  
	Any pulse shaping system is constrained by the available bandwidth which sets a limit on the minimal possible pulse duration. This limit, described using a duration-bandwidth product, states that for a given bandwidth the minimal pulse duration is achieved when the spectral phase of the pulse is at most linear \cite{milonni2010laser}. Such a pulse is termed Fourier-transform-limited. However, band-limited signals can actually oscillate locally at an arbitrarily fast rate (concomitant with a decreased amplitude), thus breaking the Fourier-transform limit resolution-wise. This is achieved through an interference phenomenon now known as superoscillation \cite{MVBerry1994}. Superoscillatory functions are known since the early 1920 when attempts were made to produce antennas with extremely narrow radiation patterns \cite{zheludev2008}.
	Such functions were revived with the introduction of quantum weak measurements \cite{aharonov1988result} which can yield values much larger than the largest eigenvalue of an observable. Berry and Popescu have introduced superoscillations into optics, predicting their use for spatial super-resolution \cite{BerryPopescu2006} which indeed was demonstrated experimentally in several works \cite{FMHuang2009, Zheludev2012, wong2013optical, tang2015ultrabroadband}. Superoscillatory diffraction-free beams \cite{Makris2011, Segev2013} as well as a superoscillating pattern made of accelerating Airy beams were demonstrated \cite{eliezer2016super}. Superoscillations were also used to realize super-narrow optical frequency conversion \cite{Remez15}. 
	In the temporal regime, relatively little has been done with attempts to break the temporal resolution limit. Two experiments done in 2006 and in 2005 used a quadratic spectral phase to break the Fourier resolution limit by $20\%$ \cite{Binhammer2006} and $30\%$ \cite{boyko2005temporal} respectively. A theoretical work discussed breaking  the Fourier limit by super-resolution pulse compression techniques \cite{Liu06}. In the radio-frequency regime temporal superoscillations were successfully demonstrated in 2011 \cite{wong2011temporal} and in 2012 \cite{wong2012superoscillatory}. In addition it was suggested that optical temporal superoscillations can be used to overcome absorption in dielectric materials\cite{Eliezer14}. 
	Here we experimentally break the temporal Fourier-transform limit of an ultra-short optical pulse by synthesizing a superoscillating pulse-envelope. In particular we achieve a temporal feature which is approximately four times narrower than a Fourier limited Gaussian pulse having the same bandwidth, while maintaining a visibility (ratio between the amplitudes of the narrow feature and its adjacent fringes) of $29.5\%$.  Numerical simulations further demonstrate the ability of such signals to achieve super-resolution in the time domain.  
	
	
	
		\emph{Theory}. We start with a known complex superoscillating function \cite{BerryPopescu2006}:
		
		\begin{equation}
			f_{SO}\left( t \right) = {\left[ {\cos \left(\Omega_0  t \right) + {i}a\sin \left(\Omega_0  t \right)} \right]^N},  \ a>1, \ N \in {\mathbb{N}}^{+}
			\label{eq:fsox}
		\end{equation}
		
		whose highest Fourier component is $N\Omega_0$, while around $t \approx 0$ it superoscillates $a$ times faster, at the rate of $aN\Omega_0$. 
		

		We expand the real part of Eq. \ref{eq:fsox} into a cosine series:
		\begin{eqnarray}
			&&\mathrm{Re} \{ f_{SO(a,N)}\left( t \right) \} = \frac{1}{2^{N-1}} \sum\limits_{k \in Even}^N {{a^k} {N\choose k}} \times \nonumber \\ \label{eq:RealSO}
			&&\sum\limits_{l=0}^{N-k} { \sum\limits_{m=0}^k { \left(-1\right)^m {N-k\choose l} {k\choose m} e^{ i  \left[ 2\left(l+m\right)-N \right] \Omega_0  t }} } \nonumber \\
			&&= \sum_{n=0}^{M=\lfloor \frac{N}{2} \rfloor} { A_{q_n} \cos { \left( q_n \Omega_0 t \right) } } \\ 
			&&q_{n} \equiv 2n + \mu_{N} \;\;\;;\;\;\; \mu_{N} \equiv mod(N,2)
		\end{eqnarray}
		
		to derive an exact set of modal real-valued amplitudes $A_{q_n}$ and frequencies creating a real-valued superoscillatory signal. Here $mod$ is the modulo operation. 
	
		A set of optical carrier modes with such amplitudes and frequencies would produce a temporal superoscillatory signal, which is quite a challenge as all the modes need to be phase locked as well as harmonics of a given fundamental frequency.
		
		It is much easier to create a superoscillation in the envelope of a given pulse which is naturally made of phase-locked modes. An interference between two modes of slightly different frequencies produces a modulated envelope beating at the difference frequency of the two modes. Interference of several beat frequencies suited with the right amplitude ratios and phases would manifest a Super-Oscillating-Beat (SOB) - an envelope with a temporal feature which breaks the temporal Fourier focusing limit.
		
		For the following, we construct a SOB signal by first setting in Eq. \ref{eq:RealSO} the parameters $N=3$ ($M=1$) and $a=2$ which gives:
%
		\begin{equation}	
			\mathrm{Re} \{ f_{SO(2,3)}(t)\} = -\frac{9}{8}\cos\left(\Omega_0 t\right) + \frac{13}{8}\cos\left(3\Omega_0 t\right)
			\label{eq:fso23}
		\end{equation}			
		
		This signal superoscillates at a rate of $6\Omega_0$. 		
		Next, the two cosine modes of Eq. \ref{eq:fso23} are  mounted on a signal's envelope according to a procedure outlined in the Supplemental material (section I) which maps each Fourier component to a beat constructed by two close-by frequency components resulting in the following modes' amplitudes and phases: 
		$|A_{q_k}|=\{ 13/8, 9/8, 9/8, 13/8 \}$, $\phi_{q_k}=\{ 0, \pi, \pi, 0\}$, $q_k=\{-3,-1,+1,+3\}$. The beats spectral spacing $\Delta\omega $ is chosen such that $(2M+\mu_N)\Delta\omega$ fits within the available bandwidth.  The frequency and time domain theoretical representations of this SOB signal are shown in the supplemental material (section I) and in Fig.\ref{fig:figure3meas}(a) below (in dashed lines).
		
	
		\emph{Results}. Our experimental setup consists of a Ti:Sapphire femtosecond laser oscillator together with a home built 4f pulse shaper and a home-built Frequency-Resolved-Optical Gating (FROG) apparatus \cite{trebino1997measuring} used for pulse characterization (see methods section in the Supplemental material). 
		Generally the SOB signal is prone to dispersion destructing the superoscillation after propagating a dispersion length of $\left({4\pi^2}\right)/\left({GVD \times (N\Delta \omega/2)^2}\right)$ where $GVD={\partial^2 k}/{\partial \omega^2} |_{\omega=\omega_0}$ and $N\Delta \omega/2$ is the bandwidth of the pulse. For the signals that we used, with bandwidth around $8$ nm,  the dispersion length in air (in BBO crystal) is in the order of one kilometer (centimeters) which is much longer than the optical path length we used (thickness of the crystal used in the FROG apparatus). Thus the distortions caused by dispersion could be ignored.
		
		The experiment went as follows: first, The pulse shaper was used to shape the original spectrum into a Gaussian and a rectangular shape, both with a flat spectral phase. The rectangle full width and the Gaussian full-width at half maximum (FWHM) are  $15$ nm ($7$ THz).
		
		In the time domain, the transform limited Gaussian pulse feature has a FWHM of $140 \pm 5$ fs  while the transform limited Sinc pulse
		has the primary and secondary lobes FWHM of $169 \pm 5$ fs and $93 \pm 5$ fs correspondingly.
		These numbers are within $97\%$  ($90\%$) of FWHM of ideal theoretical waveforms for the rectangular (Gaussian) due to experimental imperfections in the waveform synthesis. 
		The spectrogram (time-frequency FROG traces), retrieved spectrum and the temporal waveform of the Gaussian and Sinc pulses are shown in Fig. \ref{fig:figure2meas}.(a) and Fig. \ref{fig:figure2meas}.
		(b) respectively.
		
		\begin{figure*}[htbp]
			\centerline{\includegraphics[width=0.9\linewidth]{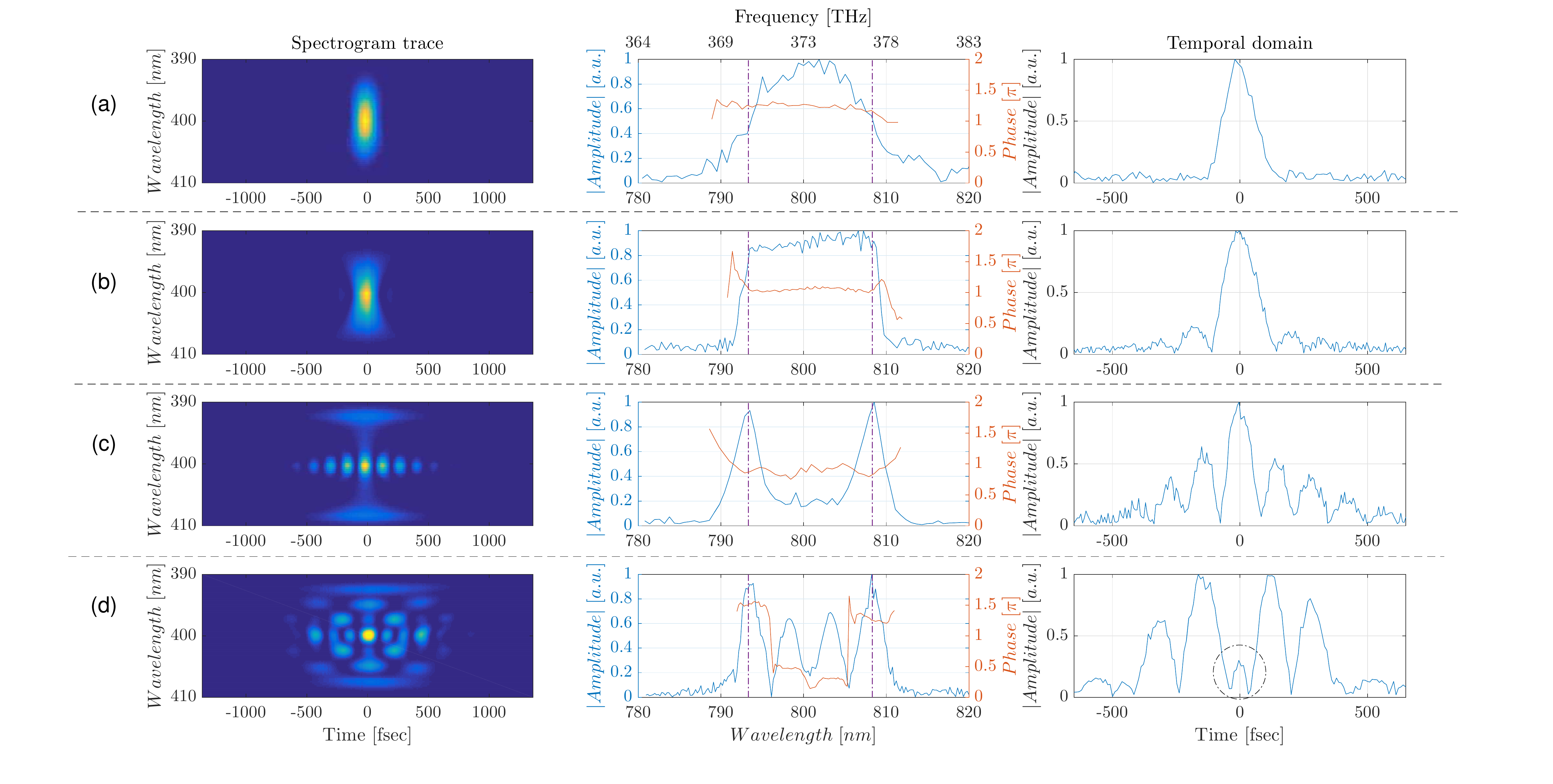}}			
			\caption{ 
				\textbf{Transform-limited signals and a super-oscillating beat.} Experimental measurements of SHG FROG time-frequency spectrograms (left column) and the retrieved signals in the frequency (middle column) and time (right column) domains for four different signals sharing the same bandwidth (which is delimited by purple dot-dashed lines): (a) transform-limited Gaussian pulse (b) transform limited Sinc pulse (c) highest frequency beat signal (d) super-oscillating optical beat (SOB). The superoscillation is circled with a dot-dashed line. 
			}
			\label{fig:figure2meas}
		\end{figure*} 	
		
		Next, the pulse shaper is set to generate a single beat comprised of two modes set $15$ nm apart (Fig. \ref{fig:figure2meas}.(c)). This is a manifestation of the fastest possible single Fourier component within the given bandwidth and it is generally assumed that it gives the narrowest possible temporal features in the form of interference fringes. The fringes FWHM is $112 \pm 5$ fs (which is off by $12\%$ of the corresponding perfect waveform). This FWHM value is  $80\%$ of the Gaussian pulse's FWHM and $66\%$ of the Sinc pulse's central lobe FWHM. This is a very general and known result: the resolution available by a (spatial or temporal) double-slit interference is better than possible with a single slit diffraction whose width is equivalent to the double-slit separation. This fact also came to fame with the introduction of Ramsey-fringes in atom interferometry \cite{ramsey1950molecular}. 
		
		Finally we synthesize the SOB signal given above for which we set the following: 
		modes amplitudes are $|A_{\{-3,-1,+1,+3\}}| = \{ {13}/{8}, {9}/{8}, {9}/{8}, {13}/{8} \} \times A_0 \;$ ($A_0$ is a common amplitude factor,); center frequencies of the modes are:  $ 
		\nu_{\{-3,-1,+1,+3\}} = {\omega_{\{-3,-1,+1,+3\}}}/({2\pi}) =  \{370.89 ,\; 373.22,\; 375.56,\; 377.89 \} \ THz$; The frequency difference between adjacent modes is $
		\Delta\nu = 2.334 \; THz \;\; (\Delta \lambda \approx $5 nm$)$ and the modes phases are $\phi_{\{-3,-1,+1,+3\}} = \{ 0, \pi, \pi, 0 \} $.

		The modes possess an approximate Gaussian form whose width $\Delta\nu$ is inversely proportional to an overall $0.8 ps$ pulse duration. The SOB spectrogram, frequency and temporal waveforms are shown in (Fig. \ref{fig:figure2meas}.(d)). It is evident that around time zero a superoscillating feature emerges, with a FWHM of $48 \pm 4$ fs (the half-maximum value was taken between the peak maximum and its adjacent minima). The SOB FWHM is approximately twice as narrow as the fringes of the corresponding single beat (double-slit) pattern, three times narrower compared with the transform limited Gaussian pulse and $3.5$ times narrower than the central lobe of the transform-limited Sinc pulse.

		Note that although both the SOB and single beat signals have spectral content extending beyond the designated spectral width of $15$ nm (due to the finite bandwidth of each mode),  it does not make the oscillating features \emph{within} the envelope narrower, and so it is irrelevant to our result.  This excess spectral content only limits the overall duration of the entire signal. 
		
		A comparison between the theoretical ideal SOB signal and the one that was synthesized can be seen in Fig. \ref{fig:figure3meas}(a) where the agreement is quite good, especially for the superoscillating feature. 
		The $87 \pm 5$ fs temporal full-width delimiting the synthesized superoscillation corresponds to a $5.75$ THz local frequency. This local frequency differs by $18\%$ from the theoretical value of 
		$a \times N \times \Delta\nu/2 = 7$ THz, which is twice the corresponding single beat frequency (which by itself corresponds to the fastest Fourier component in the SOB spectrum). The visibility of this SOB is $29.5\%$. 
		
		Because superoscillations are an interference phenomena they rely on keeping the correct amplitude and phases of their constituting modes. Still, there is some resilience to changes. For example, we have decreased the phase difference between the beam modes by $0.2\pi$ and measured the resulting temporal shape (see Fig. \ref{fig:figure3meas}(b)). In this case the measured FWHM is $72 \pm 5$ fs which is wider by approximately $50\%$ than the full width of the unmodified SOB signal.
		We note that a theoretical treatment for the sensitivity of superoscillating signals to amplitude and phase changes was made in Ref.\cite{Eliezer14} .

		An illuminating case is equalizing the modes' phases which completely ruins the superoscillation (Fig. \ref{fig:figure3meas}(c)). Here the spectral phase is linear, resulting in a transform limited pulse for which  the overall root-mean-square width is minimized \cite{weiner2000femtosecond}. Thus a linear spectral phase minimizes a global feature of the pulse - its overall width (which is also the case for the examples shown in \ref{fig:figure2meas}(a)-(c)). In contrast, when a super-oscillating function is constructed - the spectral phase is no longer linear - thus the overall width of the pulse is not minimized. What we gain, however, is a fringe (or several fringes) which is (are) narrower than the fringes of a transform limited pulse. In view of this, for super-oscillatory functions, the optimization, instead of being global is a local optimization: narrowing a given fringe while keeping the magnitude of its surrounding side-lobes as low as possible.

		\begin{figure*}[htbp]
			\centerline{\includegraphics[width=1.0\linewidth]{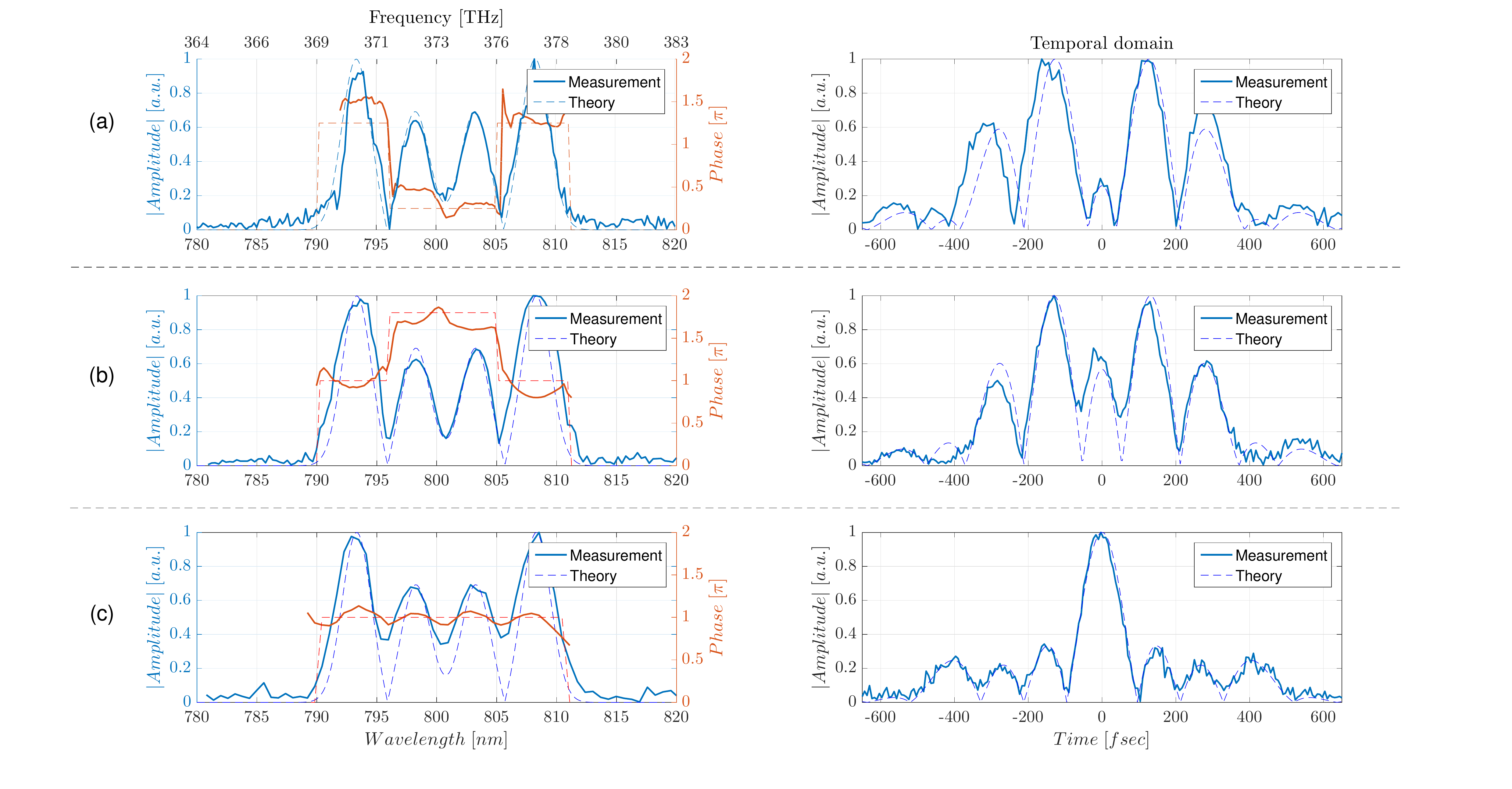}}
			\caption{ 
				\textbf{Phase modifications of a SOB signal.} Experimentally retrieved waveforms (solid lines) vs. optimal theoretical waveforms (dashed lines) in the frequency domain (Left) and the time domain (right) for three different instances of a SOB signal:
				(a) original SOB signal (b) modified by lowering the phase difference between beat modes by $0.2\pi$ (c) modified to have a flattened spectral phase. 
			}
			\label{fig:figure3meas}
		\end{figure*} 
		
		For SOB signals, the increase in local frequency comes at the expense of increased side lobes resulting in decreased visibility. For spatial super-resolution the existence of large side lobes sets fundamental limits on the resolving power of the optical system \cite{sales1997fundamental}. However, this limitation becomes irrelevant when the narrow feature interacts with an isolated small enough object, or when the side lobes can be cut by the application of a nonlinear filter. In microscopy this means the use of a pupil close to the scanned object. The pinhole projects the superoscillation into a real Fourier component. Thus the super resolving spot becomes evanescent but without the side-lobes. We would gain resolution (compared to illuminating the pinhole with a plane wave) when the superoscillating spot is smaller than the diameter of the pinhole, while the pinhole still cuts the side lobes. Similarly for the time domain - the effects of the side-lobes can be mitigated when interacting with isolated short events, or when an additional temporal gating is used. Another interesting possibility would be the use of pre-processing  for repeated applications of the SOB waveform, while changing one of its parameters, for isolating the effect of the superoscillating feature (see e.g. Ref.\cite{eramo2011method}). 
		
		Regarding the above mentioned tradeoff it is theoretically possible to continuously tune the SOB waveform between better temporal focusing to better visibility of the superoscillating feature. Most simply this is done by changing $a$ in Eq. \ref{eq:fso} which sets the ratio between the superoscillating frequency to the highest frequency in the spectrum of the signal. This is shown in Fig. \ref{fig:figure4meas}(a) where keeping the same number of modes (with N=3) and their spectral width while continuously changing $a$ between 1 to 6 results in gradually increased temporal focusing and decreased visibility. We realized experimentally three instances of the SOB with different values of $a=\{1.63, 2, 2.5\}$. The SOB with $a=2$ was shown already in Fig. \ref{fig:figure2meas} and Fig. \ref{fig:figure3meas} where the FWHM of the superoscillation was $48 \pm 5$ fs and the visibility was $29.5\%$. Fig. \ref{fig:figure4meas}(b) (Fig. \ref{fig:figure4meas}(c)) shows a SOB signal with $a=1.63$ ($a=2.5$) where the superoscillation FWHM and visibility are $78 \pm 5$  fs ($45 \pm 4$  fs) and $41\%$ ($16.8\%$) respectively.
		
		\begin{figure*}[htbp]
			\centerline{\includegraphics[width=0.95\linewidth]{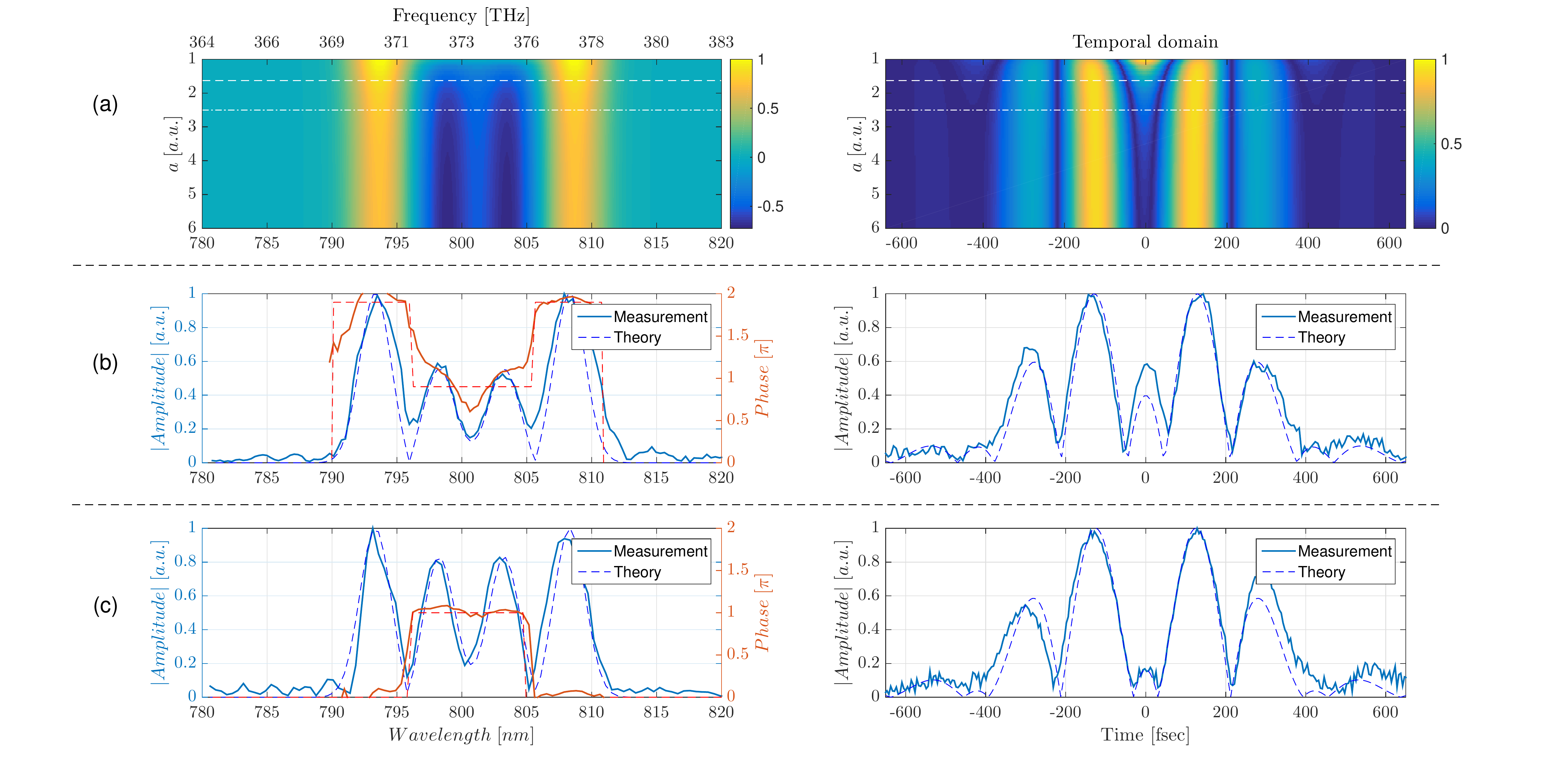}}
			\caption{ 
				\textbf{Tuning the SOB signal between better resolution to better visibility.} 
				(a) Numerical modification of the $a$ parameter in the Frequency domain (left column) and in the Temporal domain (right column). The increase in $a$ results in better temporal resolution of the SOB signal at the expense of visibility. Notice that the super-oscillating portion of the waveform is around time zero. The dashed white lines indicates the $a$ values for which waveforms where experimentally retrieved 
				(solid lines) and compared with optimal theoretical waveforms (dashed lines) as shown in the frequency domain (Left) and the time domain (right) for: (b) $a=1.63$ (c) $a=2.5$.
			}
			\label{fig:figure4meas}
		\end{figure*}
		
		Practically, when considering pulse shaping, to get a narrower SOB feature, better resolution and control is needed in the spectral domain to allow for synthesizing the required waveform. In addition, in this work, we have chosen to work with a specific family of superoscillatory functions described with 
		Eq. \ref{eq:fso}. Alternatively it is possible to work with other superoscillatory functions which optimize the duration and amplitude of the superoscillating features 
		\cite{katzav2013yield}. 

		Despite the obvious limitations mentioned above of superoscillatory wave-functions, several recent works already proved experimentally that in microscopy such waveforms can outperform transform limited beams, achieving super-resolution \cite{wong2013optical, Zheludev2012}. Due to the analogy that exists between optical phenomena in the time domain to that in the spatial domain it is expected that having a superoscillatory temporal signal would enable to achieve temporal super-resolution. Such an analogy is used in simulations presented in the Supplemental material (section II) which demonstrate numerically temporal super resolution.


	To conclude we have applied the concept of superoscillations to the temporal domain of ultra short optical pulses. We experimentally demonstrated a superoscillating optical beat having a temporal fringe which is three times narrower than a Gaussian pulse whose FWHM equals its full bandwidth, breaking the temporal Fourier-transform limit given with transform limited Gaussian pulses by $67\%$ while maintaining visibility of $29.5\%$. Such sub-Fourier focusing could be used for temporal super-resolution and so have important consequences in applications relying on ultra-short pulses such as spectroscopy, nonlinear optics and metrology.  

\newpage

	\section{Supplemental material - Constructing a super-oscillating beat (SOB) signal from a superoscillatory signal}
	
	Considering the following family of complex superoscillating functions:
	\begin{equation}
		f_{SO}\left( t \right) = {\left[ {\cos \left(\Omega_0  t \right) + {i}a\sin \left(\Omega_0  t \right)} \right]^N}, \quad \ a>1, \quad N \in {\mathbb{N}}^{+}
		\label{eq:fso}
	\end{equation}
	
	It is possible to expand the real part of Eq. \ref{eq:fso} into the following binomial expansion and Fourier cosine series:
	\begin{eqnarray}
		&&\mathrm{Re} \{ f_{SO}\left( t \right) \} = \frac{1}{2^{N-1}} \sum\limits_{k \in Even}^N {{a^k} {N\choose k}} \times \nonumber \\
		&&\sum\limits_{l=0}^{N-k} { \sum\limits_{m=0}^k { \left(-1\right)^m {N-k\choose l} {k\choose m} e^{ i  \left[ 2\left(l+m\right)-N \right] \Omega_0  t }} } \nonumber \\ 
		&&= \sum_{n=0}^{M=\lfloor \frac{N}{2} \rfloor} { A_{q_n} \cos { \left( q_n \Omega_0 t \right) } } 
		\label{eq:ftransform} 
		\\
		&&q_{n} \equiv 2n + \mu_{N} \;\;\;;\;\;\; \mu_{N} \equiv mod(N,2)
	\end{eqnarray}
	
	The SOB signal is the sum of $M+1$ beats, where those who have a beat frequency different than zero are composed of two modes having the same amplitude and phase: 
	\begin{eqnarray}	
		&&f_{SOB}(t) = (1-\mu_N) B_0 \cos(\omega_0 t + \phi_0) + \\ &&\sum_{m=-M-\mu_N}^{M+\mu_N} (1-\delta_{m,0}\mu_{N}) B_m \cos \left( \omega_m t + \phi_m\right) \nonumber \\
		&\;& B_m = B_{-m}, \;;\; \phi_m = \phi_{-m}, \;;\; \omega_0 \equiv \omega_c \\
		&=& 2(1-\mu_N) B_0 \cos(\omega_c t + \phi_0) + \nonumber \\ 
		&& 2\sum_{m=1}^{M+\mu_N} B_{m} \cos \left(\left(\frac{\omega_{-m}+\omega_{m}}{2}\right)t + \frac{\phi_{-m}+\phi_{m}}{2} \right) \times \nonumber \\
		&&\cos \left(\left(\frac{\omega_{-m}-\omega_{m}}{2}\right)t  \right) 
		\label{eq:fsob1}
	\end{eqnarray}		
	
	Here $\delta_{i,j}$ is the Kronecker delta function and $\omega_c$ is a carrier frequency within the bandwidth of the pulse.  In addition the $B_m$ amplitudes are positive valued. The beats are chosen to have the same mean frequency, while their beat frequencies are integer multiplication of an arbitrary fundamental beat frequency:
	\begin{eqnarray}	
		\forall m: \; \frac{\omega_m+\omega_{-m}}{2} &=& {\omega_c} \nonumber
		\\ 
		\frac{\omega_m-\omega_{-m}}{2} &=&  \left(\frac{2m-\mu_N}{2}\right)\Delta\omega
		\label{eq:fsobcond}
	\end{eqnarray}				
	
	These reduce Eq. \ref{eq:fsob1} to:
	\begin{eqnarray}	
		&&f_{SOB}(t) = 2\cos(\omega_c t + \phi_0) \times \\ 
		&&\left( (1-\mu_N) B_0 + \sum_{m=1}^{M+\mu_N} B_{m} \cos \left(\left(\frac{2m-\mu_N}{2}\right)  \Delta\omega t + 
		{\phi_{m}} \right) \right) \nonumber
		\label{eq:fsob11}
	\end{eqnarray}			
	
	Here $\cos \left({\omega_c} t +\phi_0\right)$ is the common carrier signal of the beats that oscillates at the frequencies $\left(({2m-\mu_N})/{2}\right) \Delta\omega$.
	Provided that the beats' amplitudes and phases are determined by Eq. \ref{eq:ftransform} 
	i.e. $B_m \equiv |A_{q_{m-\mu_N}}| \;\;;\;\; \phi_m = ({\pi}/{2})  (1-sgn(A_{q_{m-\mu_N}}))$ (where $sgn$ indicates the sign function),	
	the envelope of $f_{SOB}(t)$ would be superoscillating. While the highest beat frequency of the envelope is bounded by $\left({2M+\mu_N}/{2}\right)\Delta\omega$, still the envelope would locally superoscillate at a higher beat frequency of $a\left({2M+\mu_N}/{2}\right)\Delta\omega$.
	
	In practice, the modes constituting the SOB would have some spectral width, inducing a finite envelope width $\sigma_t$ for the temporal SOB signal while essentially not modifying the superoscillating frequency. In this case the SOB signal and its spectrum are given by:
	\small	
	\begin{eqnarray}
		&&f_{SOB}(t) = 
		2 \exp{\left( -\frac{t^2}{2\sigma_t^2} \right)} \cos(\omega_c t + \phi_0) \times 
		\label{eq:fsobFourier}						
		\\ 
		&&\left[ (1-\mu_N) |A_{q_0}| + 
		\sum_{1-\mu_N}^{M} |A_{q_m}| \cos \left(\left(\frac{2m+\mu_N}{2}\right) \Delta\omega t + 
		{\phi_{q_m}} \right) \right] \nonumber
		\\
		\label{eq:fsobFreqDomain}			
		&&F_{SOB}(\omega>0) = \frac{\sqrt{2\pi\sigma_t^2}}{2} \times 
		\\ 
		&& \sum_{-M-\mu_N}^{M} |A_{q_k}| \exp{\left( -\frac{1}{2} \sigma_t^2\left(\omega - \left[\omega_c + \left(\frac{2k+\mu_N}{2}\right)\Delta\omega\right]\right)^2 +i{\phi_{q_k}} \right)} \nonumber
		\\
		&& A_{q_k}=A_{-q_k} \;\;,\;\; \phi_{q_k}=\phi_{-q_k} 
		\nonumber
	\end{eqnarray}		
	\normalsize
	For the main text we have constructed a SOB signal by first setting in Eq. 2 the parameters $N=3$ and $a=2$ which gives:
	
	%
	\begin{equation}	
		\mathrm{Re} \{ f_{SO(2,3)}(t)\} = -\frac{9}{8}\cos\left(\Omega_0 t\right) + \frac{13}{8}\cos\left(3\Omega_0 t\right)
		\label{eq:fso23}
	\end{equation}			
	
	The Fourier representation for positive frequencies of this signal is depicted in Fig\ref{fig:figure0theory}.(a) while its time domain form is given in Fig \ref{fig:figure0theory}.(b). 
	Together with this function we also depict a cosine oscillating at the highest Fourier component of the signal $3\Omega_0$, and a cosine oscillating at the superoscillation frequency $6\Omega_0$. It is apparent that the signal superoscillates around time zero. 
	
	Then, the two cosine modes of Eq. \ref{eq:fso23} are  mounted on a signal's envelope according to the procedure outlined above which results in the following modes' amplitudes and phases: 
	$|A_{q_k}|=\{ 13/8, 9/8, 9/8, 13/8 \}$, $\phi_{q_k}=\{ 0, \pi, \pi, 0\}$, $q_k=\{-3,-1,+1,+3\}$. $\Delta\omega $ and $\omega_c$ are chosen such that $(2M+\mu_N)\Delta\omega$ fits within the available bandwidth and $\omega_c$ is the designated carrier frequency. Thus the SOB has been generated. The frequency and time domain representations of this SOB signal are shown in Fig \ref{fig:figure0theory}.(c) and Fig \ref{fig:figure0theory}.(d) respectively.

	\begin{figure*}[htbp]
		\centerline{\includegraphics[width=0.9\linewidth]{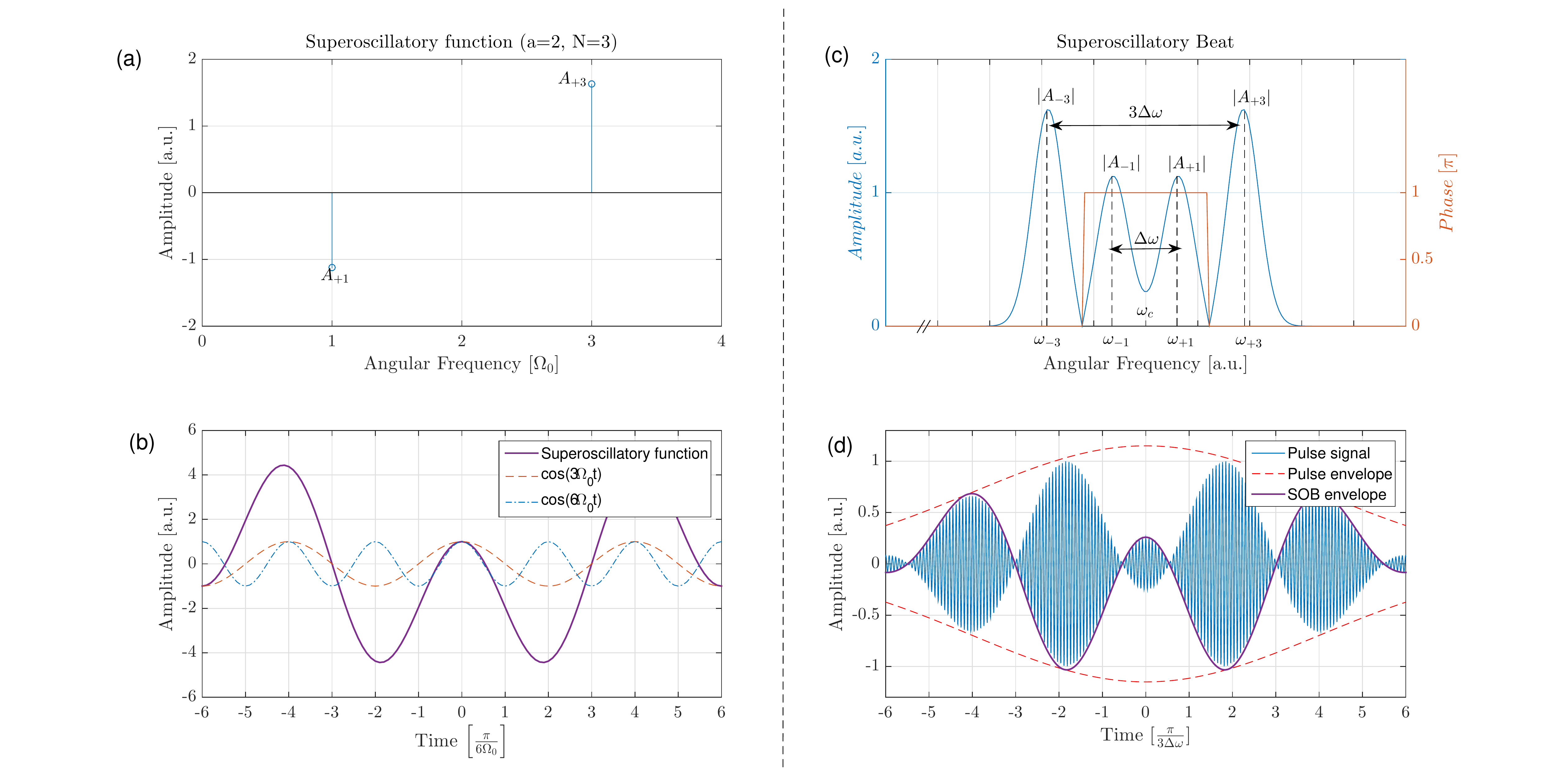}}			
		\caption{ 
			\textbf{Construction of a superoscillatory optical beat.}
			A real valued superoscillatory (SO) function is first defined through its Fourier modes which are harmonic multiples of a fundamental frequency. These modes are then reflected around a central carrier frequency to generate the superoscillating-optical-beat (SOB): a superposition of beat frequencies with a superoscillatory envelope function. (a) The positive frequency components of  the SO signal (b) Temporal waveform of the SO signal (thick continuous purple line) together with its fastest Fourier component (dashed red line) and the Fourier component corresponding to the superoscillation (dot-dashed blue line). (c) The positive frequency components of the SOB signal. (d) The temporal waveform of the SOB (continuous blue line) together with a trace of the superoscillating envelope (thick continuous purple line) and the pulse finite envelope (dashed red line) which is due to the finite width of the constituting Fourier modes. 
		}
		\label{fig:figure0theory}
	\end{figure*}
	
	\section{ Temporal super-resolution with SOB signals }
	
	Here we bring the results of numerical simulations applying an analogy with microscopy to demonstrate temporal super-resolution using a SOB signal.  The analogy with the spatial case is quite straightforward. A spatial imaging system is described through the convolution of a point-spread-function and the object to be imaged. With a superoscillating point-spread-function super-resolution is achieved \cite{wong2013optical}. In our case - the physical signal to be used in a generic measurement would be an optical polarization vector proportional to the mixing of the SOB signal with a temporal event signal $g(t)$: $P \propto f_{SOB}(t)g(t-\tau)$. Here $\tau$ is the relative delay between the two real-valued signals. If we further assume that the overall interaction length is short, then a slow intensity detector would measure the cross-correlation signal $S(\tau)=\int \left[f_{SOB}(t)g(t-\tau)\right]^2 dt$. 
	We wish to analyze the detection of a temporal double-peak modeled as two separate Gaussian pulses: 
	\begin{equation}	
		g(t) = \left(
		e^{ -\frac{(t-\frac{1}{2}t_{sep})^2}{2\sigma^2} } +
		e^{ -\frac{(t+\frac{1}{2}t_{sep})^2}{2\sigma^2} } 
		\right) \cos(\omega_gt)
	\end{equation}			
	
	with a carrier frequency $\omega_g$. For the following we fix $\sigma=0.15 \times t_{sep}$. In the simulations we set the carrier frequencies of both the SOB signal and $g(t)$ to zero to factor out the fast oscillations associated with the carrier frequency of the polarization (formally this is equivalent to the application of a low-pass filter to the cross-correlation). We numerically calculated the cross-correlation for various values of $t_{sep}$ and for various SOB signals by modifying the $a$ parameter for two values of the $N$ parameter: $N=3,4$. The SOB signals are normalized by their energy. In Fig. \ref{fig:figure5theory}(a) we see two examples of SOB signals with $N=3,4$ and with $a=3.4, 3.25$ (correspondingly) superimposed with a temporal double peak signal $g(t)$ with some small separation $t_{sep}$. In Fig. \ref{fig:figure5theory}(b) the cross-correlations are given separately for $N=3,4$ for a specific value of $t_{sep}= 0.32 \times \left[{2\pi}/{N\Delta\omega}\right]$ while $a$ is modified. The cross-correlations are shown only around time-delay-zero where the superoscillating feature is interacting with the double-pulse. The curved white lines delimits the range  $\tau \in [-T_{SOB}/4, T_{SOB}/4]$ (where $T_{SOB}={4\pi}/({aN\Delta\omega}$) which reflects the temporal delays in which the double pulse interactions with the lobes outside the superoscillatory feature is minimal. The delay between the two straight vertical white lines is equal to $t_{sep}$.
	The two-pulse structure is resolved when a minima occurs at time-zero of the cross-correlation (for a single pulse we would get a maxima at this location). However the resolving power is really a matter of visibility  - how well is this feature observable. We calculate the visibility of the central feature of the cross-correlation (not to be confused with the visibility of the superoscillating feature) for different values of $a$, where the cross-correlation visibility is $|({max(S)-min(S)})/({max(S)+min(S)})|$. The maxima and minima are calculated for the range  $\tau \in [-T_{SOB}/4, T_{SOB}/4]$ . The $a$ value where the visibility is maximal is denoted with a horizontal straight white line. The greatest visibility is achieved for $a$ for which the two-pulse separation  matches the distance between the closest zeros of the superoscillation feature. This condition is approximately given by: $T_{SOB}/2=t_{sep}$.  This happens for both $N=3,4$. Furthermore - when we repeat the calculation of the visibility for different values of $t_{sep}$ we still see this identical behavior. This is shown in Fig. \ref{fig:figure5theory}(c) depicting the visibility as a function of both $t_{sep}$ and $a$. The maximal visibility approximately matches the line $t_{sep} = T_{SOB}/2$ (shown with black dots).  
	The explanation for the fact that there is an optimal value of $a$ for resolving the double-pulse is simple: it is the result of  trade-off between higher local frequency associated with higher values of $a$ and lower visibility due to lower ratio of the amplitude of the superoscillations compared to its adjacent side-lobes.  
	In any case, the conclusion is obvious: if the double-pulse separation is shorter than half the period associated with diffraction limited signals, than a superoscillating signal would be better for detecting or resolving it (compared with a transform-limited pulse for which $a=1$), achieving super-resolution in the  time-domain.	
	
	We would like to add that the temporal resolving power  is a function of the signal to be resolved. In analogy with imaging -  regular microscopes are defined usually by their Modulation Transfer Function (MTF) which shows the visibility when imaging a specific Fourier component. The MTF is irrelevant for microscopes based on super-oscillations, as the power of the later lies in their ability to resolve signals made of a limited number of oscillations (not a Fourier component). If the number of oscillations extends too much into the side-lobes - they would not be resolved. In our case this reflects the cases where $t_{sep}$ is fixed and $a$ is increased too much. As we have seen the performance of the SOB signals would outperform transform-limited signals for cases where the signal to be resolved does not extend into the side-lobes of the SOB signal. This is the temporal counterpart to super-resolution imaging demonstrated experimentally with super-oscillating microscopes \cite{wong2013optical, Zheludev2012}. 
	
	\begin{figure*}[htbp]
		\centerline{\includegraphics[width=0.95\linewidth]{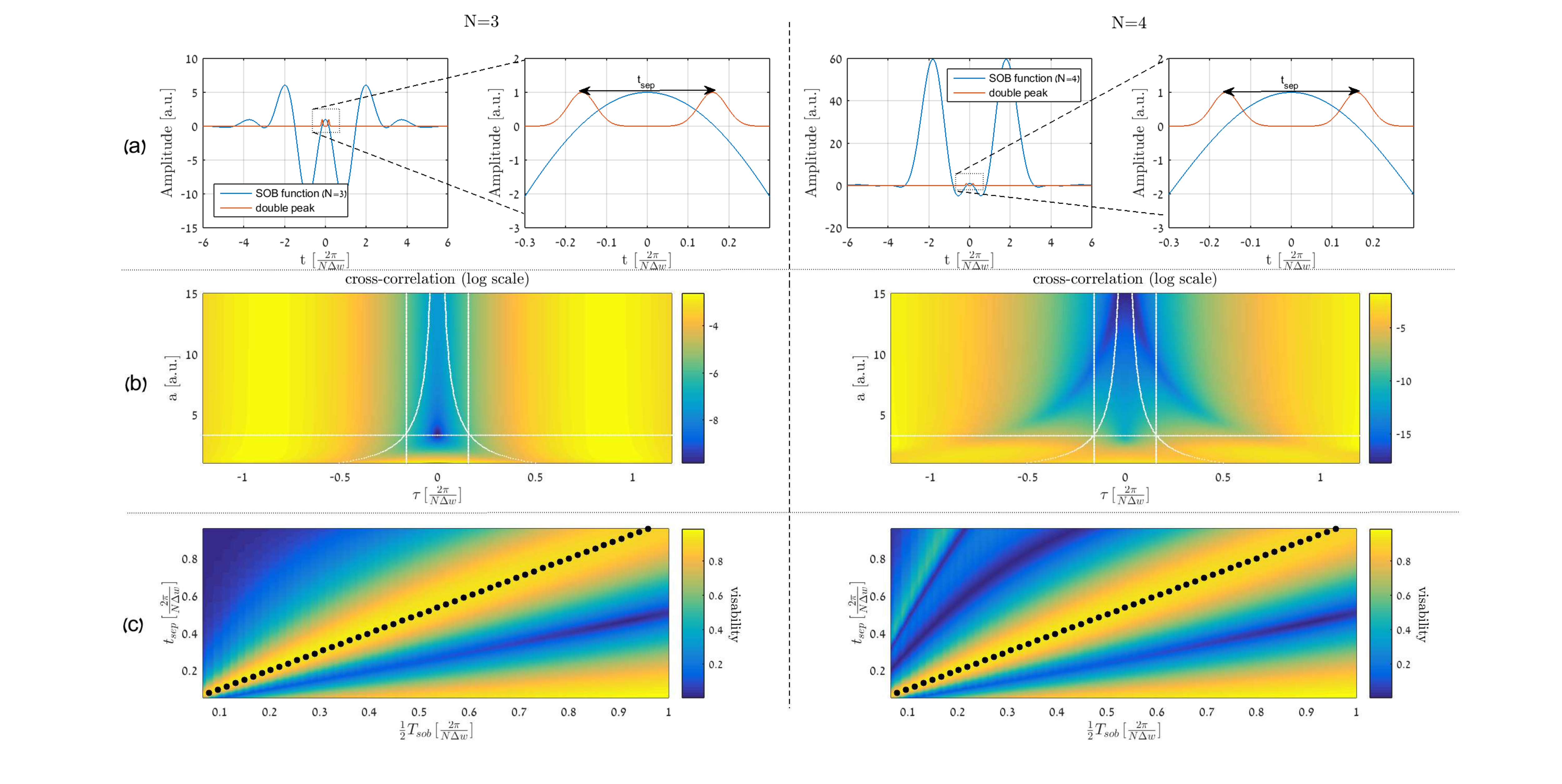}}
		\caption{
			\textbf{Temporal super-resolution with SOB signals}
			(a)  SOB signals (blue line) with $N=3, a=3.4$ (left) $N=4, a=3.25$ (right)  superimposed with a temporal double peak signal $g(t)$ (red) with some small separation. 
			(b) Cross-correlation function of the SOB signal together with a double-peak signal with a specific separation $t_{sep}=0.32 \times ({2\pi})/({N\Delta\omega})$, given separately for $N=3$ (left) $N=4$ (right) as a function of time delay and the $a$ parameter. The separation of the two vertical straight lines is $t_{sep}$. The horizontal white line marks the $a$ value for which the visibility of the cross-correlation is maximal in the range $\tau \in [-T_{SOB}/4, T_{SOB}/4]$. This range is delimited between the two curved white lines.
			(c) Visibility as a function of $t_{sep}$ and $T_{SOB}/2={1}/{a}$ (in the units used in the graph) over the range  $\tau \in [-T_{SOB}/4, T_{SOB}/4]$. The maximal visibility approximately matches the line $t_{sep} = T_{SOB}/2$ (shown with black dots). 
		}
		\label{fig:figure5theory}
	\end{figure*}
	
	\section{Methods}
	
	In our experiment, we use a home-made Frequency Resolved Optical Gating (FROG) apparatus and a home-made  Pulse Shaper.
	
	The FROG was built using a $50:50$ beam splitter, a $50 \mu m$ BBO SHG crystal, a $0.1 \mu m$ step linear motor stage,
	and an off-axis parabolic mirror having a reflected focal length of  $4''$ .
	
	The pulse shaper was built using a pair of $35cm$ focal length cylindrical mirrors,
	and a pair of 1200 ${lines}/{mm}$ holographic gratings.
	At the Fourier plane we used a 640 pixel, dual-mask Spatial Light Modulator (Jenoptik SLM-S640d).
	The laser source used in the experiments was Coherent Vitara-T.
	
	Fig. \ref{fig:figure1setup} depicts a detailed schematic of our experimental setup.
	
	\begin{figure}[ht]
		\centerline{\includegraphics[width=1.0\linewidth]{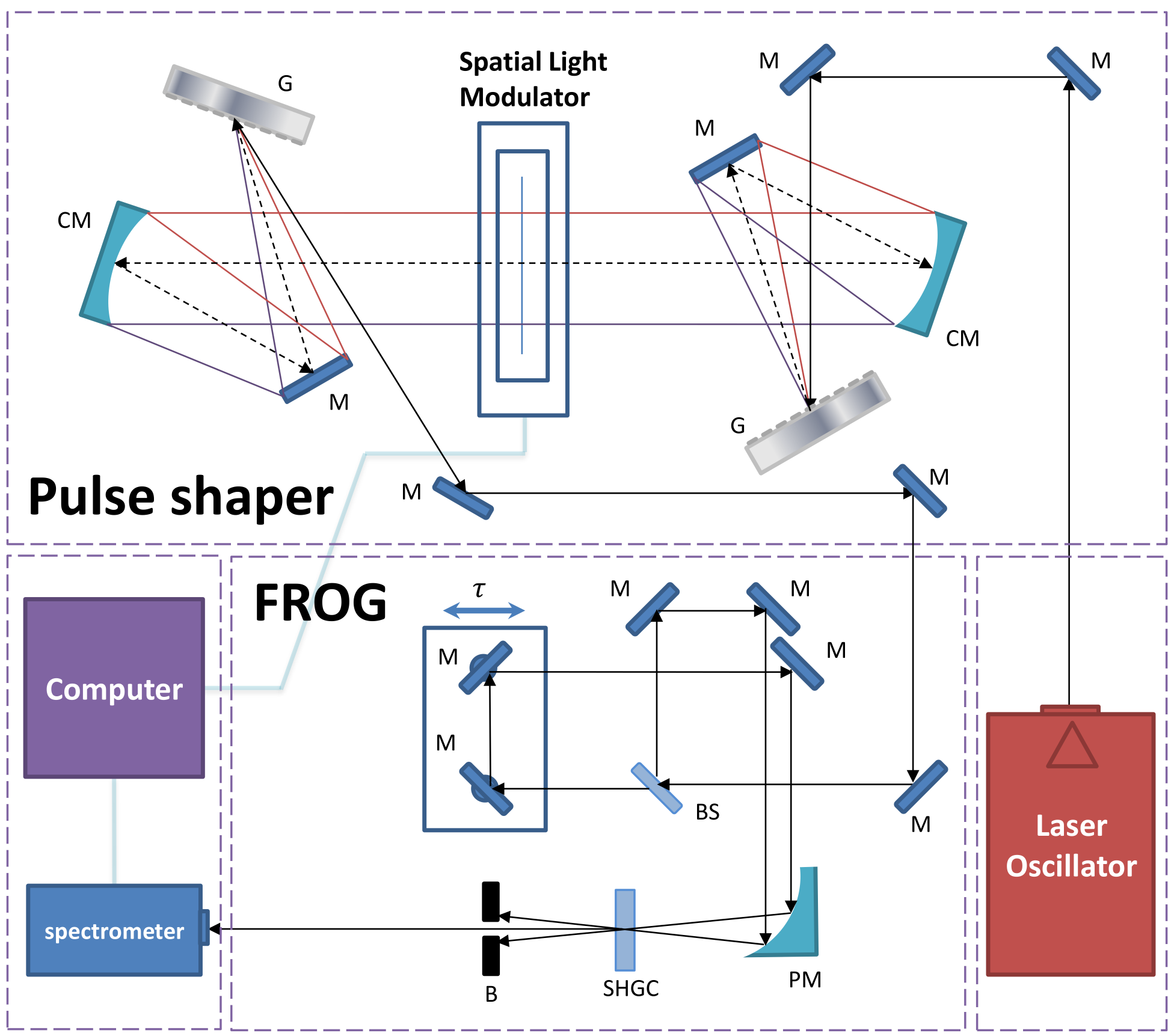}}
		\caption{
			Experimental setup. The pulses emitted by an ultra-fast laser oscillator are shaped in a 4f Fourier domain pulse shaper. The shaped pulses amplitude and phase are retrieved through a measurement in a FROG apparatus. M=Mirror, CM=Cylindrical Mirror, G=Grating, BS=beam Splitter, PM=Off-axis Parabolic Mirror, SHGC=Second-Harmonic-Generation Crystal, B=Beam Blocker.  
		}
		\label{fig:figure1setup}
	\end{figure}
	
	All FWHM measurements were done using a 2nd order polynomial fit over the retrieved waveforms.
	Indicated uncertainties in experimentally retrieved values  are based on the temporal and spectral resolution of our FROG apparatus.

%



\bibliography{bibfile} 

\end{document}